\documentclass[12pt,preprint]{aastex}

\begin{document}

\title{Magellan PFS Discovery of a Cold Gas Giant Exoplanet Likely Formed by a Gas Disk 
Gravitational Instability}

\author{Alan P.~Boss$^1$, R. Paul Butler$^1$, Johanna K. Teske$^{1,2}$, Guang-Yao Xiao$^3$, 
Yifan Xuan$^3$, Fabo Feng$^3$, Jiayin Li$^4$, Sharon X. Wang$^4$, Stephen A. Shectman$^2$, \& 
Jeffrey D. Crane$^2$}

\affil{$^1$Carnegie Science, Earth \& Planets Laboratory, 5241 Broad Branch Road NW, 
Washington, DC 20015-1305, USA}
\affil{$^2$The Observatories of the Carnegie Institution for Science,
813 Santa Barbara St., Pasadena, CA 91101, USA}
\affil{$^3$State Key Laboratory of Dark Matter Physics, Tsung-Dao Lee Institute \& School of Physics 
and Astronomy, Shanghai Jiao Tong University, Shanghai 201210, People’s Republic of China}
\affil{$^4$Department of Astronomy, Tsinghua University, Beijing 100084,
People's Republic of China}

\email{aboss@carnegiescience.edu}

\begin{abstract}

 Using the Planet Finder Spectrograph (PFS) on the 6.5-m Magellan Clay Telescope at the Las Campanas
Observatory we have discovered a cold gas giant exoplanet orbiting the K7V dwarf star HIP88962. 16 
yrs of PFS observations imply an exoplanet with an orbital period of $\approx$ 19.3 yrs and a radial
velocity semiamplitude (K) of 94.9 m s$^{-1}$. With a host star mass of 0.62 $M_\odot$, this implies a 
planet mass $M_p sin(i) \sim 5 M_{Jup}$.
While stable M dwarfs have a PFS RMS of $\approx$ 3 m s$^{-1}$, the single Keplerian fit for
HIP88962 has an RMS of $\approx$ 6 m s$^{-1}$, hinting at a second exoplanet. The best double
planet Keplerian fit, with a 4.113 d period, still results in a RMS of $\approx$ 5 m s$^{-1}$.
HIP88962 has Ca II H lines with twice the emission of four stable stars, implying that
stellar activity may contribute to the RMS jitter error for the single Keplerian fit. When
combined with Hipparcos and Gaia astrometry, HIP88962 b has a nearly edge-on orbit with inclination 
91 $\pm$ 22 deg, true planet mass $4.83$ M$_{Jup}$, semimajor axis 6.15 $\pm$ 0.5 au,
and eccentricity 0.718 $\pm$ 0.02. Including a PFS jitter of $\approx$ 7 m s$^{-1}$
yields $\chi^2 = 0.995$ for the combined solution. A massive gas giant orbiting at 
$\approx$ 6 au around a 0.62 $M_\odot$ star is incompatible with published 
core accretion population synthesis models, but consistent with formation by gas disk
gravitational instability.

\end{abstract}

\section{Introduction}

 There are 21 known G stars within 10 pc of the sun, but at least 236 M dwarfs 
(T. J. Henry et al. 1997; A. Reiners \& G. Basri 2009). M dwarfs have twice
as many planets as G dwarfs with orbits inside $\sim$ 1 au (G. D. Mulders et al. 2015).
Red dwarfs are thus attractive targets for finding the closest habitable worlds 
(e.g., C. D. Dressing \& D. Charbonneau 2015), those most 
amenable to follow-up observations. M stars are estimated to harbor about 37\% of the 
habitable zone real estate within 10 pc of the Sun (J. R. Cantrell et al. 2013). Estimates 
of the frequency of habitable Earths around M dwarfs are as high as 53\% (R. K. Kopparapu 2013). 
M dwarfs tend to have super-Earths and sub-Neptunes rather than gas giant companions
(e.g., S. Sabotta et al. 2021), with super-earths outnumbering the sub-Neptunes for 
orbital periods less than 7 days around mid-to-late M dwarfs (K. Ment \& D. Charbonneau 2023).
E. K. Pass et al. (2023) found no Jupiter-mass companions inside the ice lines in their 
radial velocity (RV) survey of 200 mid-to-late M dwarfs.
However, surveys such as the GEMS survey are now detecting increasing
numbers of super-Jupiters orbiting M dwarfs (S. Kanodia et al. 2024; A. Hotnisky et al. 2025).
M dwarf exoplanet demographics beyond the ice lines ($\sim$ 0.03 to $\sim$ 0.3 au;
periods of $\sim$ 10 to $\sim$ 100 days) are still poorly understood.

 The PFS M Dwarf Survey (MDS), which 
began in 2010, focuses on relatively bright, early M dwarfs (e.g., M1 to M4) in order to minimize the 
exposure time needed to undertake a blind search for exoplanets, as opposed to determining masses for 
known TESS exoplanet detections (e.g., J. K. Teske et al. 2021) where longer exposures 
on fainter target stars are worthwhile in order to measure the planet's mass. 
G. Anglada-Escud\'e et al. (2013) used PFS data to show that the M1.5V dwarf GJ 667 C is orbited
by at least six exoplanets, including three super-Earths in its habitable zone. Nearby K
dwarfs have been studied as well. P. Arriagada et al. (2013) detected two exoplanets with minimum masses
of $\approx 53 M_\oplus$ and $\approx 6.5 M_\oplus$ orbiting the K7 dwarf GJ 221 with orbital
periods of 125 d and 3.87 d, respectively.

 The MDS target list includes several other late K dwarfs (K5 to K7), one of which (HIP88962) is the 
subject of the present paper. HIP88962 is a possible (Tier 3) target for the Habitable Worlds Observatory 
(N. Tuchow et al. 2025), raising the importance of its planetary system. Previous radial velocity 
studies of HIP88962 found no evidence of Doppler variations (L. A. Paredes et al. 2021) using the 
CHIRON spectrograph on the CTIO/SMARTS 1.5-m telescope. Their 10 observations were limited to
a time span of 799 d, insufficient to detect the Doppler wobble discovered by the MDS. Here we present 
the PFS discovery of a long period, massive gas giant around HIP88962 that appears unlikely to have
been formed by the classical core accretion mechanism for gas giant planet formation
(H. Mizuno 1980) or by its more recent variants. We shall see that gas disk gravitational instability 
(GDGI; A. P. Boss 1997) is a more promising mechanism for the formation of HIP88962 b.

\section{Data \& Star}

 The Doppler data was taken with the Planet Finder Spectrograph (PFS; J. D. Crane et al. 2006, 2008, 
2010) on the 6.5-m Magellan Clay telescope at the Las Campanas Observatory in Chile. 47 epochs
of data on HIP88962 were taken with PFS on 41 nights between 2010 and 2025.
PFS currently has a resolution of 130,000 and covers wavelengths from 388 nm to 668 nm with
high precision and stability. PFS uses an iodine cell (G. W. Marcy \& R. P. Butler 1992) as a 
wavelength reference to enable accurate measurement of the wavelength of the host star's spectral 
lines as they are periodically Doppler shifted by the orbital variations of the star's radial
velocity around the center of mass of the star and planetary system (R. P. Butler et al. 1996).

 The astrometric data was obtained by the Hipparcos (M. A. C. Perryman 1997) and Gaia (Gaia 
Collaboration et al. 2017) space satellites, launched by the European Space Agency in 1989 and 
2013, respectively. Gaia provides microarcsecond astrometry for over $10^9$ stars, enabling the 
proper motion and wobble of a target star around the system center of mass to be determined.
Astrometry enables the orbital inclination ($i$) of the exoplanet orbit to be measured, 
whereas Doppler velocities are sensitive only to the radial component of the stellar wobble
and so yield only a lower limit ($M_p sin(i)$) on an exoplanet's true mass ($M_p$). 
F. Feng et al.(2023) and G.-Y. Xiao et al. (2024) present the means used to combine the 
Doppler and astrometric data to determine $i$ and hence $M_p$ (see also Y. Wu et al. 2026).
The combination of radial velocities with astrometry to determine true exoplanet masses
is a powerful means for comparing observed exoplanet population demographics with the predictions 
of exoplanet masses by various theoretical models of planet formation (e.g., K. Cui et al. 2026).

 HIP88962 (GJ 4042) is a K7V dwarf with a V magnitude of 10.47 and B-V index of 1.38 at a 
distance of 27.3 pc. We adopt a stellar mass of $0.620 \pm 0.084\,M_\odot$ for HIP88962 
from TIC v8.0/v8.2 (Stassun et al. 2019). HIP88962 is not included in the TIC Cool Dwarf List; 
instead, its mass was inferred using the standard TIC $T_{\rm eff}$-mass spline relation. 
Specifically, its Gaia DR2 color $G_{\rm BP}-G_{\rm RP}=1.7769$, with zero reddening adopted 
because of its proximity, gives $T_{\rm eff}=3993 \pm 122 K$, which was converted to the 
reported mass using the relation in Table 3 of Stassun et al. (2019).

 For comparison with the PFS data for HIP88962, we also present PFS velocities for four stars with 
similar B-V indices which do not show clear evidence for a Doppler wobble produced by an exoplanet.
These four stable stars  are: HIP105090 (HD202560), HIP38594, HIP66675 (HD118926), and
HIP42601, with V magnitudes of 6.69, 9.75, 9.61, and 10.53, respectively, B-V indices
of 1.397, 1.385, 1.376, and 1.360, respectively, and RMS velocities of 2.70 m s$^{-1}$,
2.94 m s$^{-1}$, 2.99 m s$^{-1}$, and 3.48 m s$^{-1}$, again respectively.

\section{Results}

\subsection{Radial Velocity (RV) Analysis}

 Figure 1 shows the best single (top) planet solution and the fit residuals (bottom) for the PFS data on 
HIP88962. The single planet fit yields an orbital period of $\approx$ 19 yrs 
with K = 94.9 m s$^{-1}$, planet mass $M_p sin(i) \sim 5 M_{Jup}$, and orbital eccentricity $\sim 0.7$. 
The large residuals in Figure 1 suggest that at least one more exoplanet might be present. 
The single planet orbit residuals suggest a second planet with a period of 4.11 days
on a circular orbit. However, the second planet fit residuals are little improved over the single 
planet fit, suggesting that stellar jitter might be the actual source of the RMS error in the
single planet Keplerian. The 48 unbinned PFS velocities used in Figure 1 are attached to the
electronic version of this paper as a machine readable file. 

 Figure 2 shows the PFS data for four stable PFS stars with similar B-V indices as
HIP88962, showing that the velocity RMS of these four stable stars are all $\sim$ 3 m s$^{-1}$,
This result then supports 
the idea that another possibility for the residuals to the single Keplerian fit is stellar activity. 
Figure 3 shows the Ca II H line for HIP88962 and the four comparison stars. The H line emission is 
similar for all four of these stars, but HIP88962 (black) has roughly twice the H line emission of 
the stable stars. For this reason it is possible that the excess RMS of the single planet Keplerian 
fit can be attributed to stellar activity.

 We further examined the $\sim$ 4.1-day periodicity in the residuals after subtracting the long-period 
Keplerian model using generalized Lomb--Scargle (GLS) periodograms of the RV residuals, the Ca II S-index, 
and the H$\alpha$ index, together with the spectral window function. The RV residuals show their 
strongest peak at $P \simeq 4.11$ d, with a global false-alarm probability (FAP) of approximately $3\%$, 
estimated from 10,000 permutation trials. Although this indicates that the signal is unlikely to be 
produced by white noise alone, a low FAP does not by itself imply a planetary origin. In addition to the 4.11-day 
peak, the RV residuals contain prominent power near 5.85, 7.30, and 8.80 d. These timescales are particularly 
noteworthy in light of the $\sim$ 17-day stellar rotation period inferred from the TESS photometry, since 
several of them lie close to the expected rotational harmonics $P_{\rm rot}/2$, $P_{\rm rot}/3$, and 
$P_{\rm rot}/4$. Neither the H$\alpha$ nor the S-index periodogram show significant periodicity,
with no feature in their power periodograms rising as high as even a 0.1 false alarm probability. 
The spectral window further indicates that 
some of the secondary periodogram structure may be influenced by the temporal sampling.
Therefore we do not consider the 4.1-day signal to be a robust planet candidate.

 We also examined the long-term behavior of the activity indicators measured from the PFS spectra 
(Figure 4). Of particular interest is the 2010--2015 interval, during which the RVs exhibit a statistically 
significant downward trend of $-2.78\pm0.82~{\rm m\,s^{-1}\,yr^{-1}}$ ($p=0.003$). In contrast, neither 
the Ca~II S-index nor the H$\alpha$ index shows a statistically significant temporal trend over the same 
interval ($p=0.15$ and $0.13$, respectively). The contemporaneous RVs are also not significantly correlated 
with either activity indicator (Pearson $r= -0.255$, $p=0.253$ for the S-index and $r=0.185$, $p=0.423$ for 
H$\alpha$). The quoted $p$-values are from two-sided tests of the corresponding null hypotheses of zero 
slope or zero Pearson correlation. As an additional check, after subtracting the best-fitting Keplerian model, 
the residual RVs show no significant correlation with either activity indicator and no significant long-term 
trend over 2010--2015. We therefore find no evidence that a long-term stellar activity cycle is responsible 
for the 2010--2015 RV decline. Nevertheless, the enhanced Ca~II emission suggests that stellar activity 
may contribute to the short-timescale RV scatter.

 We also examined the TESS photometry of HIP88962. Rather than using the default Pre-search Data 
Conditioning (PDC) light curve, which may attenuate variability on long timescales, we retrieved the TESS 
Target Pixel Files (G. R. Ricker et al. 2015) using \texttt{lightkurve} (Lightkurve et al. 2018) and analyzed 
the pixel-level photometry with the Systematics-Insensitive Periodogram (SIP; C. Hedges et al. 2020) method.
The Sector~80 light curve of HIP88962 exhibits clear low-frequency photometric modulation, whereas 
the quiet comparison star HIP105090 shows substantially weaker variability when analyzed in the same manner. 
The periodogram of HIP88962 displays a broad dominant peak centered at $P \simeq 16.9$ d. Given the limited 
duration of a single TESS sector, which covers only about 1.5 cycles at this period, we interpret this as 
an approximate rotational timescale of $\sim 17$ d rather than a precise rotation-period measurement.
Together with the enhanced Ca II H emission, the stronger photometric variability of HIP88962 provides 
independent evidence for appreciable stellar activity. We therefore consider stellar activity to be 
a plausible contributor to the excess RV scatter about the single-Keplerian model.

 Finally, we have also considered additional RV models: one Keplerian plus a linear trend, and one 
Keplerian plus linear and quadratic trends. The results for these three models are RMS values 
of 6.525, 6.26, and 5.74 m s$^{-1}$, and Bayesian information criteria (BIC) of 251.017, 251.359, and 
246.949, respectively. Although the RMS decreases, the BIC improvement is not strong ($\Delta$BIC $<$ 5). 
However, the inferred orbital periods and semimajor axes increase with the additional trends, particularly 
for the quadratic model, where the period becomes more than twice as long (17726 d), while the uncertainties 
remain comparable. Given the limited orbital coverage, we believe that it is reasonable to retain the original 
orbital solution and to avoid over-interpreting the alternative fits. Continued RV monitoring of HIP 88962 should 
help to further clarify the exoplanet's orbit.

\subsection{Joint Analysis of Radial Velocities \& Astrometry}

 Radial velocity (RV) observations constrain only the minimum planetary mass, $M_p sin(i)$, 
because the orbital inclination is generally unknown. The long temporal baseline between Hipparcos 
and Gaia astrometry observations provides sensitive astrometric constraints on stellar reflex motion, 
helping to break this degeneracy and determine the true planetary mass and orbital orientation.

 We jointly model the RV measurements together with Hipparcos (F. van Leeuwen 2007) and Gaia second 
and third data releases (GDR2 and GDR3; Gaia Collaboration et al. 2018, Gaia Collaboration et al. 2023) 
following the framework developed in our previous studies (F. Feng et al. 2023; G.-Y. Xiao et al. 2024).
Briefly, the RVs constrain the spectroscopic orbital parameters, while the astrometric model accounts 
for the barycentric motion, parallax, and the stellar reflex motion induced by the planetary companion. 
For Hipparcos, we model the intermediate astrometric data. In addition, the offset between Hipparcos 
astrometry and the astrometry propagated from the GDR3 epoch to the Hipparcos epoch are also 
incorporated. For Gaia, because epoch astrometry is not yet publicly available, we use the observing 
epochs, scan angles, and parallax factors predicted by the Gaia Observation Forecast Tool 
\footnote{\url{https://gaia.esac.esa.int/gost/index.jsp}} (GOST) to generate synthetic along-scan 
measurements. These measurements are subsequently fitted with the standard five-parameter astrometric 
model and compared with the catalog solutions from GDR2 and GDR3.

 We adopt uniform priors for most fitting parameters and derive the orbital solution by sampling 
the posterior via the parallel-tempering Markov Chain Monte Carlo (MCMC) sampler \texttt{ptemcee} 
(W. D. Vousden et al. 2016). We employ 20 temperatures, 100 walkers, and 50,000 steps per chain to 
generate posterior distributions for all the fitting parameters, with the first 40,000 steps being 
discarded as burn-in.  We assess convergence by visually inspecting the cold-chain traces and by 
computing the Gelman--Rubin statistic (A. Gelman \& D. B. Rubin 1992); all fitted parameters satisfy 
$\hat{R}<1.01$. Further details and validation of the joint RV--astrometry framework are presented in 
G.-Y. Xiao et al. (2026). Figure 5 then presents the orbit plot illustrating the nearly edge-on 
nature of this system, as derived from the Hipparcos and Gaia GDR2/GDR3 data.

 Finally, Table 1 presents the best Bayesian estimate of the system parameters for HIP88962 b as 
derived from the joint analysis of radial velocities and the Hipparcos and Gaia astrometry shown in Figure 5. 
With the combined solution, HIP88962 b has an orbit with inclination $91 \pm 22$ deg, a true planet mass 
of $4.83^{+0.63}_{-0.51}\,M_{\rm Jup}$, a semimajor axis of $6.15 \pm 0.5$ au, and an orbit with eccentricity 
$0.718 \pm 0.03$. Including a fitted PFS RV jitter term of $\approx 6.85 \ {\rm m\,s^{-1}}$ gives a reduced $\chi^2$ 
of 0.995 for the RV component.

\section{Formation Mechanisms}

 How could a 4.83 $M_{Jup}$ gas giant form around a 0.62 $M_\odot$ K7V star and end up on an
orbit with a semimajor axis of 6.15 au with an eccentricity of 0.718? 
Core accretion (H. Mizuno 1980) is the conventional method for the formation of the great
majority of gas giant exoplanets, with planetesimal collisions leading to $\sim 10 M_\oplus$
cores capable of dynamically accreting gaseous envelopes (e.g., J. B. Pollack et al. 1996).
Several variants have been proposed in order to speed up the process of the collisional 
accumulation of solids and overcome bottlenecks such as collisional erosion rather than growth.
B. Liu et al. (2019) considered the pebble accretion variant of core accretion around stars with
masses from $\sim$ 0.1 $M_\odot$ to 1 $M_\odot$. Their Figure 7 shows a dramatic decrease in
the formation of gas giants as the stellar mass is reduced below $\sim 0.5 M_\odot$, with no
gas giants forming with masses above $\approx 2 M_{Jup}$ even for 1 $M_\odot$ stars.

 R. Burn et al. (2021) presented a population synthesis model for low mass stars, finding
that planetesimal accretion was unable to form any gas giants for stars with masses
below 0.5 $M_\odot$ (their Figure 8). While their Figure 8 inferred gas giant planet formation for
a stellar mass of 0.62 $M_\odot$, their Figure 7 shows that essentially none of the planets
had masses greater than $\sim 1 M_{Jup}$ with semimajor axes greater than 1 au
for stars with masses of 0.5 $M_\odot$ or 0.7 $M_\odot$. Similarly, M. Schlecker et al. (2022)
used the same population synthesis model to show that no massive gas giants could end up on orbits
with periods greater than $10^3$ d for stars with masses of 0.5 $M_\odot$ or 0.7 $M_\odot$
(their Figure 4). For comparison, HIP88962 b's orbital period is 7055 days (Table 1). 
The R. Burn et al. (2021) models followed the dynamical interactions between the growing 
protoplanets, including mutual scattering, and found that the gas giants that did manage to form
had mean orbital eccentricities of 0.19 and 0.14 for 0.5 $M_\odot$ and 0.7 $M_\odot$ stars,
respectively. For comparison, HIP88962 b's orbital eccentricity is 0.718 (Table 1).

 The updated planet population synthesis model by planetesimal accretion of T. Kimura \& M. 
Ikoma (2026) found no planets with mass greater than $\sim 0.3 M_{Jup}$ on final orbits beyond 1 au
for 0.5 $M_\odot$ stars (their Figure 9), though a few Jupiter-mass gas giants ended up inside 
$\sim 10$ au for a 1 $M_\odot$ star (their Figure 6). 

 The alternative extreme for gas giant planet formation is the GDGI (A. P. Boss 1997). 
A. P. Boss \& S. Kanodia (2023) showed that GDGI could form gas giants around stars
with masses from 0.1 $M_\odot$ to 0.5 $M_\odot$. For a 0.5 $M_\odot$ star, their models
formed gas giants with masses ranging from $\sim 1 M_{Jup}$ to $\sim 20 M_{Jup}$
(their Figure 5), provided that the protoplanetary disk mass was at least 0.03 $M_\odot$
(their Figure 6) for a range of disk temperatures of 10 K to 60 K. 
A. P. Boss (2006) found that GDGI could lead to gas disk fragmentation into clumps
with initial orbits in the range of 4 au to 10 au for 0.5 $M_\odot$ stars.
A. P. Boss (2023) studied the orbital migration of protoplanets formed by GDGI around
1 $M_\odot$ stars, finding that mutual gravitational encounters and spiral arm 
interactions could scatter multiple Jupiter-mass planets to semimajor axes of 30 au
and beyond, and orbital eccentricities close to unity for masses below $\sim 10 M_{Jup}$.
While none of these GDGI models performed a population synthesis model on the scale
of the core accretion models previously described, it is clear that the GDGI mechanism 
is quite capable of having formed an exoplanet similar to HIP88962 b.

\section{Conclusions}

 The Magellan MDS PFS team has discovered a relatively distant, cold, massive gas giant orbiting 
the K7V dwarf star HIP88962, whose existence confounds the conventional wisdom about the
formation of gas giant planets. The discovery of HIP88962 b was not the first to confound
the formation models. A. Sozzetti (2023) used Doppler velocities and Hipparcos/Gaia
astrometry to detect a $3.6 M_{Jup}$ gas giant orbiting at 3.5 au around a 0.49 $M_\odot$
star. A. Hotnisky et al. (2025) used Doppler velocities and TESS photometry to detect
two super-Jupiters, with masses of $7.8 M_{Jup}$ and $10 M_{Jup}$ on short period
orbits, around a M0 and a M2 star, respectively. Similar to the present paper,
both A. Sozzetti (2023) and A. Hotnisky et al. (2025) interpreted their discoveries
as evidence for formation by the GDGI. Y. Wu et al. (2026) used Doppler velocities
and Hipparcos/Gaia astrometry to detect four cold Jupiters, though the host stars
all had masses in the range of 0.80 $M_\odot$ to 1.38 $M_\odot$. 
More such discoveries are certain to follow, especially from Gaia astrometry,
as we learn more about the demographics of exoplanets around low mass stars.
Tantalizing evidence for several more cold gas giants orbiting K7 dwarfs
is awaiting confirmation by future Magellan PFS observations. These detections are 
further proof of the need for a hybrid, universal model of exoplanet formation combining
both extremes: core accretion and gas disk gravitational instability (e.g., K. Cui et al. 2026).

\acknowledgments

 We thank the Magellan Clay Telescope operators along with the entire staff of the 
Las Campanas Observatory for their unflagging support over several decades of the
PFS instrument and observational programs. We also thank the reviewer for a number 
of thoughtful comments and questions that have led to the improvement of this paper,
as well as for a second review that led to a correction to our Ca II S-index and H$\alpha$
values plotted in Figure 4. Two co-authors used ChatGPT as an aid for their language editing 
and some of their analysis/code development regarding the astrometric solution, which 
they checked afterwards. This research was not supported by any U.S. federal grant funds.

\clearpage
\begin{deluxetable}{cc}
\tablecaption{Results for 48 unbinned PFS velocities combined with Hipparcos and Gaia DR2/3 astrometry
for HIP88962 b, assuming a stellar mass of 0.62 $M_\odot$. The PFS velocities used are attached to the
electronic version of this paper as a machine readable file. \label{tbl-1}}
\tablehead{\colhead{quantity} & 
\colhead{value $\pm$ uncertainty}}
\startdata
Period [day] &                                 7055 +688 -481 \\
K [m s$^{-1}$] &                                 94.9 +5.2 -4.2 \\
$e$          &                                   0.718 +0.032 -0.029 \\
omega [deg] &                                  348.4 +2.0 -2.1 \\
$i$ [deg]    &                                 91 +22 -23 \\
Omega [deg]   &                                210 +36 -33 \\
Tp [JD]       &                                2466941 +687 -479 \\
PFS RV jitter [m s$^{-1}$] &                     6.85 +0.88 -0.73 \\
mp [$M_{Jup}$] &                               4.83 +0.63 -0.51 \\
semimajor axis [au] &                          6.15 +0.46 -0.39 \\
Hipparcos jitter [mas] &                       3.5 +1.5 -1.9 \\
Gaia error inflation factor  &                 1.071 +0.073 -0.049 \\
RA offset in GDR3 [mas]      &                 1.3 +1.1 -1.5 \\
Dec offset in GDR3 [mas]      &                2.18 +0.48 -1.1 \\
parallax offset in GDR3 [mas]  &               -0.003 +0.013 -0.014 \\
proper motion offset in RA in GDR3 [mas/yr] &  -0.091 +0.071 -0.060 \\
proper motion offset in Dec in GDR3 [mas/yr] & -0.127 +0.041 -0.042 \\
Final reduced chi2 for RV          &           0.995, Num: 48, DOF: 42 \\
Final RMS for RV [m s$^{-1}$]  &                   6.525 \\

\enddata
\end{deluxetable}

\clearpage

\begin{figure}
\vspace{-2.0in}
\plotone{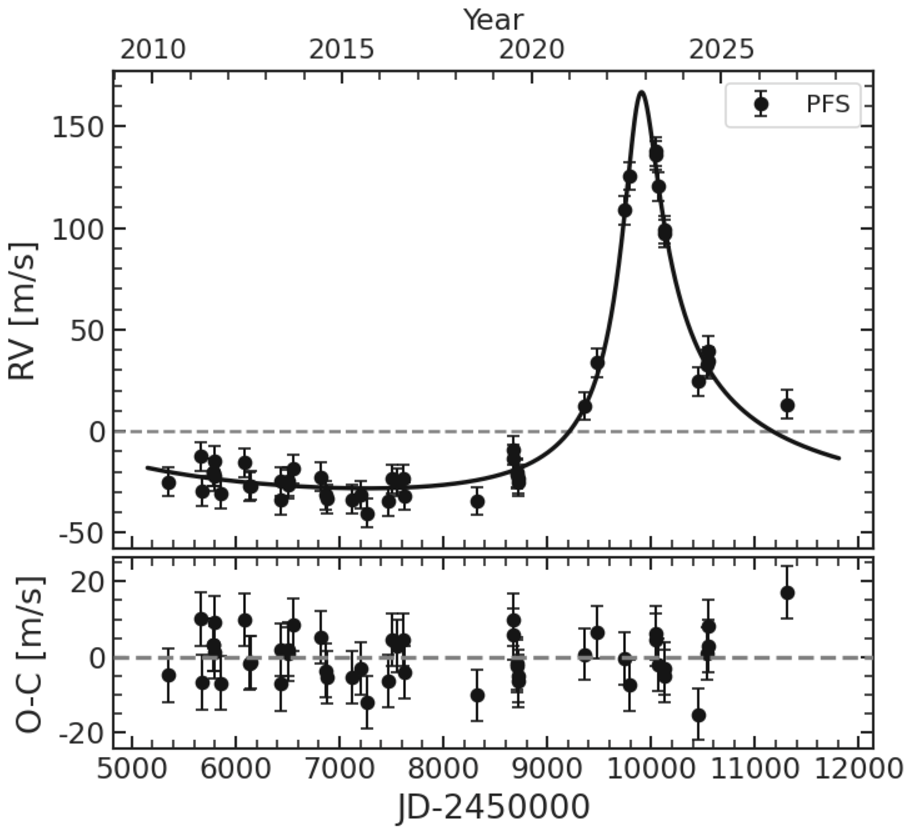}
\vspace{-2.0in}
\caption{Single (top) planet solution and residuals (bottom) for the PFS data on HIP88962.
The combined RV+astrometry single planet fit yields an orbital period of $\approx$ 19.3 yrs with 
K = 94.9 m s$^{-1}$, planet mass $M_p = 4.83 M_{Jup}$, and orbital eccentricity 0.718. 
When stellar jitter of 6.85 m s$^{-1}$ is included, $\chi^2 = 0.995$.}
\end{figure}

\clearpage

\begin{figure}
\vspace{-2.0in}
\plotone{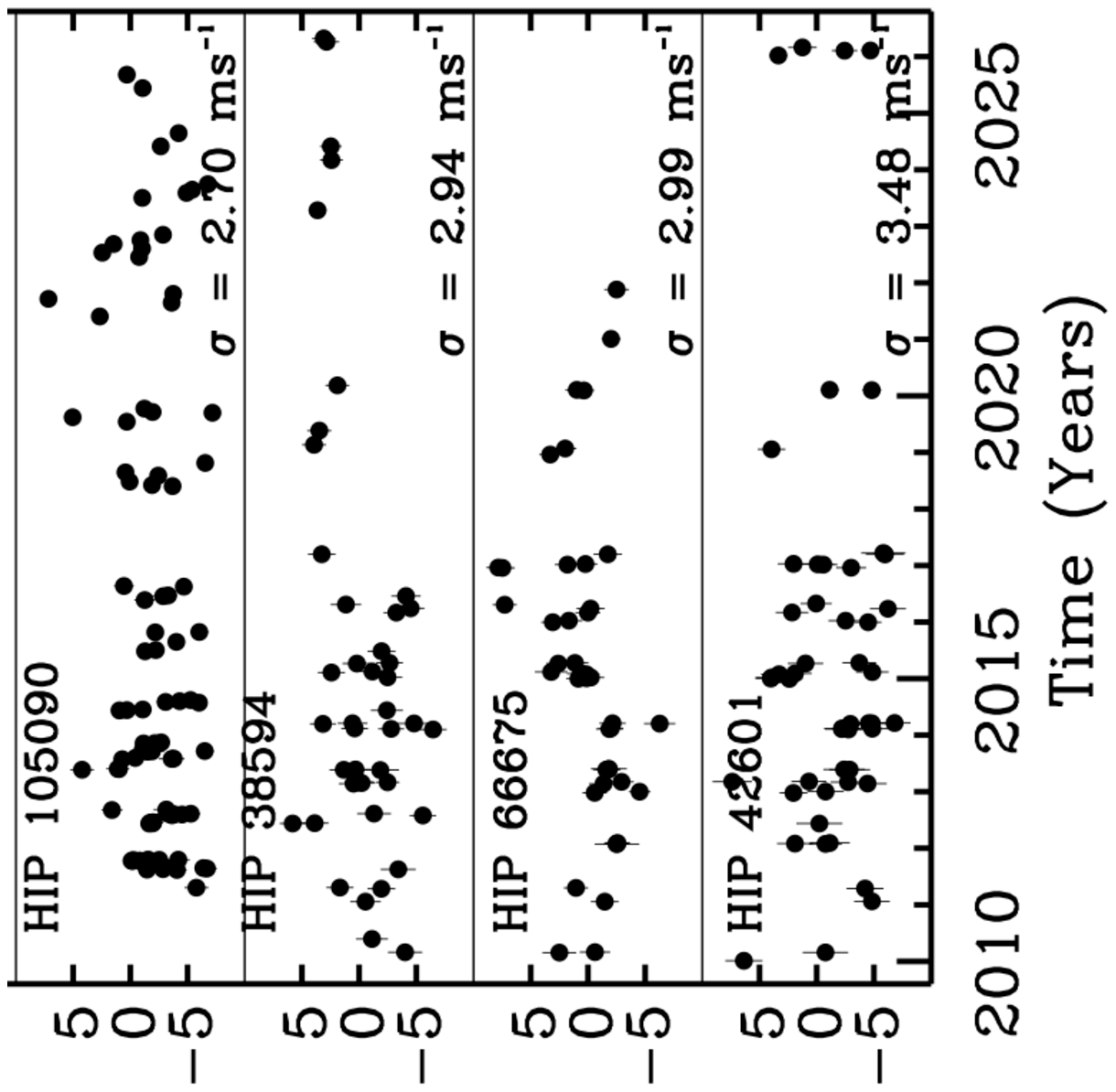}
\caption{PFS data (RV in m s$^{-1}$) for four stable M dwarfs with similar B-V indices 
to HIP88962. The RMS velocity errors of these four stable stars are all $\sim$ 3 m s$^{-1}$.}
\end{figure}

\clearpage

\begin{figure}
\vspace{-2.0in}
\plotone{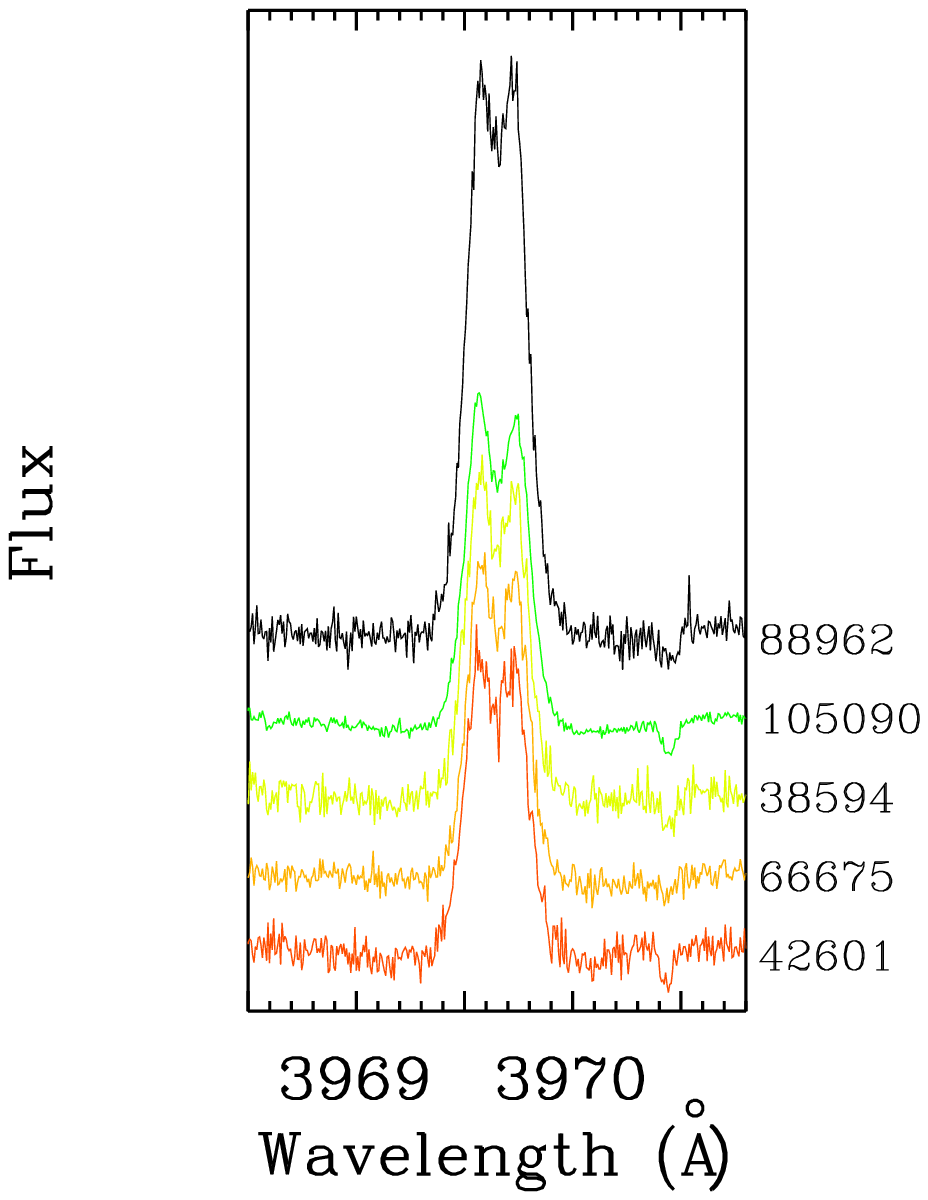}
\caption{Ca II H line fluxes for HIP88962 (black) and the four stable M dwarf stars (colors) with similar 
B-V indices. Evidently HIP88962 has a significantly higher level of stellar activity than these
stable stars.}
\end{figure}

\clearpage

\begin{figure}
\vspace{-3.0in}
\plotone{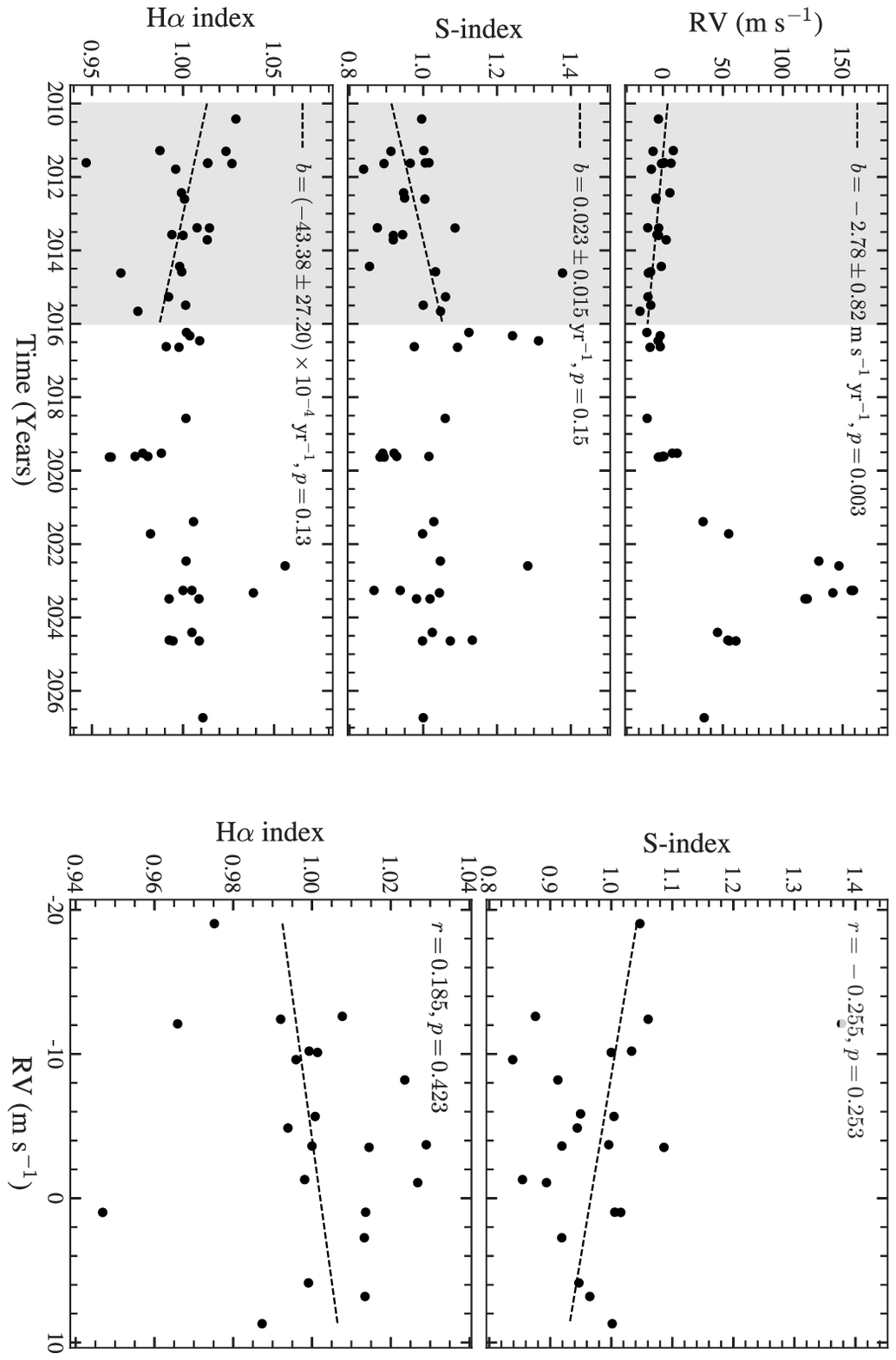}
\vspace{-0.0in}
\caption{Long-term evolution of the PFS radial velocities (top left), Ca~II S-index (middle left), 
and H$\alpha$ index (bottom left) for HIP88962, together with the correlations between the contemporaneous 
RVs and the two activity indicators during 2010--2015 (right panels). The shaded region marks the 2010--2015 
interval discussed in the text. The dashed lines in the left panels show independent linear fits over this 
interval, with the fitted slopes and corresponding $p$-values indicated. The RVs show a statistically 
significant downward trend, whereas neither activity indicator shows a significant temporal trend. The 
right panels compare the original RVs directly with the S-index and H$\alpha$ index over the same interval; 
the dashed lines show linear fits, and the Pearson correlation coefficients and corresponding $p$-values 
are indicated. Neither activity indicator shows a significant correlation with the RVs.}
\end{figure}

\clearpage

\begin{figure}
\vspace{-3.0in}
\plotone{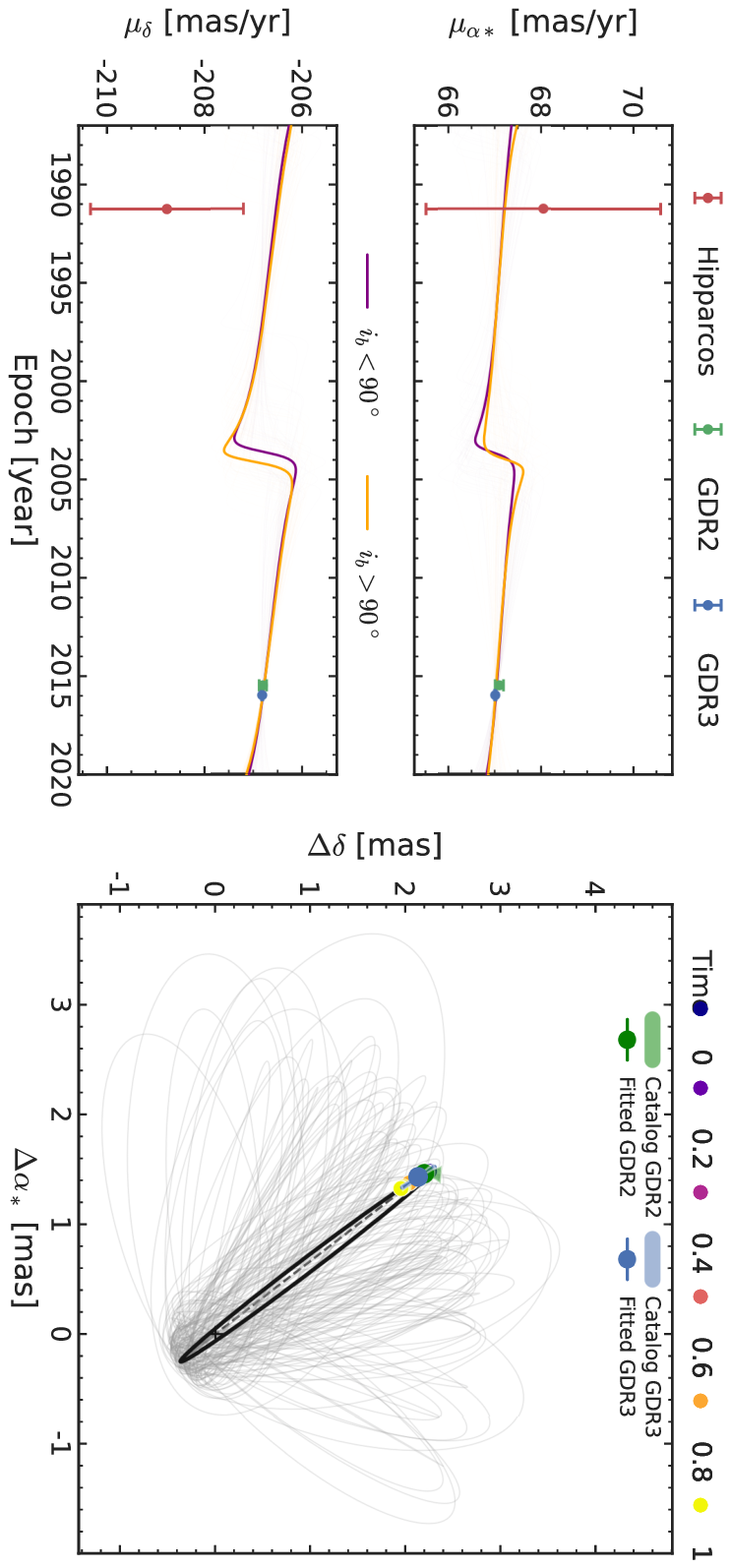}
\vspace{-1.0in}
\caption{Solution for Hipparcos-Gaia astrometry for HIP88962. Left panel: Proper-motion evolution of HIP88962 predicted by the joint fit. 
The thick purple and orange curves show the instantaneous model proper motion for the two inclination branches, $i_{b} < 90^\circ$  
and $i_{b} > 90^\circ$, respectively, both of which include the fitted barycentric proper motion and time-dependent reflex proper 
motion induced by HIP88962 b. The red, green, and blue points denote the Hipparcos, GDR2, and GDR3 catalog proper motions with their uncertainties. 
Right panel: Astrometric reflex orbit of HIP88962 induced by HIP88962 b. The plus sign marks the system barycenter. 
The thick black curve shows the best-fit orbit, and the thin gray curves show 100 MCMC posterior draws. 
The green and blue shaded regions indicate the barycentric-motion-subtracted uncertainty envelopes for GDR2 and GDR3.
The green and blue circles mark companion-induced position offsets at the GDR2 and GDR3 reference epochs, and the associated solid 
lines show their evolution due to the fitted proper-motion offsets.  The colored points along the black orbit show the model-predicted 
astrometric positions at the Gaia scanning epochs and are color-coded by normalized observing time, increasing from purple to yellow.}
\end{figure}

\clearpage

\end{document}